\documentclass[pra,fleqn]{revtex4}

\usepackage{amsmath,graphicx}
\usepackage{dcolumn}

\begin{document}

\title{Relativistic corrections of $m\alpha^6$ order to the
ro-vibrational spectrum of $\mbox{H}^+_2$ and $\mbox{HD}^+$ molecular
ions}

\author{Vladimir I. Korobov}
\affiliation{Bogolyubov Laboratory of Theoretical Physics, Joint Institute
for Nuclear Research, Dubna 141980, Russia}

\begin{abstract}
The major goal of the high-precision studies of ro-vibrational states in
the hydrogen molecular ions is to provide an alternative way for improving
the electron-to-proton mass ratio, or the atomic mass of electron. By now
the complete set of relativistic and radiative corrections have been
obtained for a wide range of ro-vibrational states of $\mbox{H}_2^+$ and
$\mbox{HD}^+$ up to order $R_\infty\alpha^4$. In this work we complete
calculations of various contributions to the $R_\infty\alpha^4$ order by
computing the relativistic corrections to the binding energy of electron.
\end{abstract}

\maketitle

In recent years several laser spectroscopy experiments have been proposed
\cite{Gremaud98,Schiller03} for high precision measurements of the
vibrational spectrum of the hydrogen molecular ions $\mbox{H}_2^+$ and
$\mbox{HD}^+$. These experiments have metrological interest and are aimed
at a sup-ppb precision. In order to improve the present accuracy of the
electron-to-proton mass ratio \cite{memp} the uncertainty of the
spectroscopic data (as well as of the theoretical calculations of the
spectra to compare to) should be below 1 part per billion(1 ppb). To meet
these stringent requirements, the theoretical calculations should achieve
at least a level of 10 kHz (or $\sim\!10^{-11}$ in atomic units).

While the variational calculations of the nonrelativistic energies have
reached a numerical precision of $10^{-15}-10^{-24}$ a.u.\ %
\cite{paris,Kor00,Bai02,Yan03,Drake04,schiller-korobov}, the radiative and
relativistic corrections have not been presented in the literature with
required accuracy. Only recently high-precision variational calculations
for the ro-vibrational states in the range of the total orbital momentum
$L\!=\!0\!-\!4$ and vibrational quantum number $v\!=\!0\!-\!4$ for the
$\mbox{H}_2^+$ and $\mbox{HD}^+$ molecular ions along with relativistic
and radiative corrections of orders $R_\infty\alpha^2$,
$R_\infty\alpha^2(m/M)$, $R_\infty\alpha^3$, $R_\infty\alpha^3(m/M)$, and,
partially, $R_\infty\alpha^4$ have been obtained \cite{KorPRA06}.

In this paper we use the CODATA02 recommended values \cite{CODATA02} for
numerical calculations, and the atomic units ($\hbar=e=m_e=1$) are
adopted.

\section{Radiative corrections of order \boldmath$R_\infty\alpha^4$.}

For a given relative accuracy of $\sim\!10^{-10}-10^{-11}$ recoil
corrections of orders $R_\infty\alpha^4(m/M)$ and higher are small and may
be neglected. That allows to reduce calculation of higher order
corrections for the Coulomb three-body system to the problem of a bound
electron in an external field.

The radiative corrections of order $R_\infty\alpha^4$ in the {\em external
field} approximation are known in an analytic form \cite{SapYen,Eides01}:
\begin{equation}
\begin{array}{@{}l}
\displaystyle
E_{se}^{(4)} =
\alpha^4
      \frac{4\pi}{m_e^2}
         \left(\frac{139}{128}-\frac{1}{2}\ln{2}\right)
         \left\langle
            Z_1^2\delta(\mathbf{r}_1)\!+\!Z_2^2\delta(\mathbf{r}_2)
         \right\rangle,
\\[4mm]\displaystyle
E_{anom}^{(4)} = \alpha^2\frac{\pi}{m_e^2}
         \left[
            \left(\frac{\alpha}{\pi}\right)^2
            \left(
               \frac{197}{144}+\frac{\pi^2}{12}-\frac{\pi^2}{2}\ln{2}
                  +\frac{3}{4}\zeta(3)
            \right)
         \right]
         \left\langle
            Z_1\delta(\mathbf{r}_1)\!+\!Z_2\delta(\mathbf{r}_2)
         \right\rangle,
\\[4mm]\displaystyle
E_{vp}^{(4)} = \frac{4\alpha^3}{3m^2}
         \left[\frac{5\pi\alpha}{64}\right]
         \left\langle
            Z_1^2\delta(\mathbf{r}_1)\!+\!Z_2^2\delta(\mathbf{r}_2)
         \right\rangle,
\\[4mm]\displaystyle
E_{2loop}^{(4)} = \frac{\alpha^4}{m_e^2\pi}
   \left[
      -\frac{6131}{1296}-\frac{49\pi^2}{108}+2\pi^2\ln{2}-3\zeta(3)
   \right]
         \left\langle
            Z_1\delta(\mathbf{r}_1)\!+\!Z_2\delta(\mathbf{r}_2)
         \right\rangle.
\end{array}
\end{equation}
The last equation includes both Dirac form factor and polarization
operator contributions.

\begin{figure}[t]
\caption{Adiabatic "effective" potentials for the relativistic $m\alpha^6$
order correction for $\mbox{H}_2^+$ molecular ion ($Z_1=Z_2=1$). Energies
are in  $\hbox{(atomic units)}\times\alpha^4$.}\label{ef_pot}
\begin{center}
\vspace{-5mm}
\includegraphics[width=70mm]{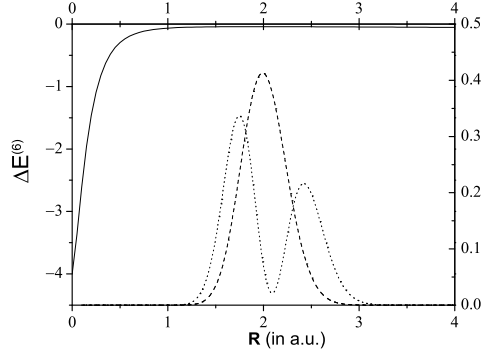}
\vspace{-5mm}
\end{center}
\end{figure}

\section{Relativistic corrections of order \boldmath$R_\infty\alpha^4$.}

The most problematic contribution of $R_\infty\alpha^4$ order  is the
relativistic correction for a Dirac electron. It can be obtain within the
adiabatic two-center approximation as follows (for details, see
\cite{KorJPB07}).

We start from the nonrelativistic  Schr\"odinger equation with the
Hamiltonian:
\begin{equation}
H_0 = \frac{p^2}{2m}+V, \qquad V=-\frac{Z_1}{r_1}-\frac{Z_2}{r_2}.
\end{equation}

The total contribution to the energy of a bound electron at the
$R_\infty\alpha^4\sim m_ec^2\alpha^6$ order is defined by
\begin{equation}\label{E4}
\Delta E^{(6)} =
   \left\langle H_B Q (E_0-H_0)^{-1} Q H_B \right\rangle
  +\left\langle H^{(6)} \right\rangle.
\end{equation}
Here $H^{(6)}$ is the effective Hamiltonian for the interaction of an
electron with the external field of two centers in this order, which can
be expressed in a form:
\begin{equation}\label{h6}
\begin{array}{@{}l}
\displaystyle
H^{(6)} = \frac{p^6}{16m^5}
     +\frac{(\boldsymbol{\mathcal{E}}_1\!+\!\boldsymbol{\mathcal{E}}_2)^2}
                                                                {8m^3}
     -\frac{3\pi}{16m^4}
        \Bigl\{
           p^2\bigl[\rho_1\!+\!\rho_2\bigr]+
           \bigl[\rho_1\!+\!\rho_2\bigr]p^2
        \Bigr\}
     +\frac{5}{128m^4}\left(p^4V\!+\!Vp^4\right)
     -\frac{5}{64m^4}\left(p^2Vp^2\right),
\\[3mm]\displaystyle\hspace{80mm}
\boldsymbol{\mathcal{E}}_i=-Z_i\mathbf{r}_i/r_i^3,
\qquad
\rho_i=Z_i\delta(\mathbf{r}_i).
\end{array}
\end{equation}
$H_B$ is the Breit--Pauli interaction:
\begin{equation}
H_B = -\frac{p^4}{8m^3}
   + \frac{\pi}{2m^2}[Z_1\delta(\mathbf{r}_1)+Z_2\delta(\mathbf{r}_2)]
   + \left(
          Z_1\frac{[\mathbf{r}_1\times\mathbf{p}]}{2m^2r_1^3}+
          Z_2\frac{[\mathbf{r}_2\times\mathbf{p}]}{2m^2r_2^3}
      \right)\mathbf{s}\>,
\end{equation}

Both terms in Eq.~(\ref{E4}) are divergent. In order to remove the
infinities a transformation to the second order term can be applied which
separates a divergent part:
\begin{equation}\label{trans}
\left\{
\begin{array}{@{}l}
H'_B = H_B+(H_0-E_0)U+U(H_0-E_0)\\[2mm]
\displaystyle
\left\langle H_B Q (E_0-H_0)^{-1} Q H_B \right\rangle =
    \left\langle H'_B Q (E_0-H_0)^{-1} Q H'_B \right\rangle
\\[2mm]\displaystyle\hspace{20mm}
    +{\left\langle UH_B\!+\!H_BU \right\rangle
    -2\left\langle U \right\rangle \left\langle H_B \right\rangle
    +\left\langle U(H_0-E_0)U \right\rangle}
\end{array}
\right.
\end{equation}
with
$
U=\frac{1}{4m}[Z_1/r_1+Z_2/r_2]=-\frac{1}{4m}V
$.

The last three terms of the second equation in (\ref{trans}) can be recast
in a form of a new effective interaction:
\begin{equation}
\begin{array}{@{}l}
\displaystyle
H'^{(6)} =
    (UH_B+H_BU)-2U\langle H_B \rangle-U(E_0-H_0)U
\\[2mm]\displaystyle\hspace{40mm}
    = \frac{p^4V\!+\!Vp^4}{32m^4}
    -\frac{\pi V\!\left[\rho_1\!+\!\rho_2\right]}{4m^3}
    +\frac{(\boldsymbol{\mathcal{E}}_1\!+\!\boldsymbol{\mathcal{E}}_2)^2}
                  {32m^3}
    +\frac{V}{2m}
                 \left\langle H_B \right\rangle.
\end{array}
\end{equation}
Here $ \rho_i=Z_i\delta(\mathbf{r}_i)$ and
$\boldsymbol{\mathcal{E}}_i=-Z_i\mathbf{r}_i/r_i^3$.

Taking into account that $ \Psi_0$ is a solution of the Schr\"odinger
equation $H_0\Psi_0 = E_0\Psi_0$, one may obtain from the above the
following finite expression \cite{KorJPB07}:
\begin{equation}
\begin{array}{@{}l}
\displaystyle
\Delta E^{(6)} =
   \left\langle H'_B Q (E_0-H_0)^{-1} Q H'_B \right\rangle
   +\Bigl\langle H^{(6)} \Bigr\rangle
   +{\Bigl\langle H'^{(6)} \Bigr\rangle}
  =
  \left\langle H'_B Q (E_0-H_0)^{-1} Q H'_B \right\rangle
\\[2mm]\displaystyle\hspace{25mm}
  +\frac{3E_0\left\langle V^2 \right\rangle}{4m^2}
  -\frac{5E_0^2\left\langle V \right\rangle}{4m^2}
  -\frac{3\pi E_0\left\langle(\rho_1+\rho_2)\right\rangle}{4m^3}
  +\frac{\left\langle\mathbf{p}V^2\mathbf{p}\right\rangle}{8m^3}
  +\frac{\left\langle V \right\rangle\left\langle H_B \right\rangle}{2m}
  +\frac{E_0^3}{2m^2}.
\end{array}
\end{equation}
This new expression along with
modified second order iteration can be now calculated numerically.

"Effective" potentials of $\Delta E^{(6)}(R)$ have been obtained for
different bond lengths in \cite{KorJPB07}. Results are shown in
Figure~\ref{ef_pot}.

Averaging them over the radial wave function of particular state one may
get corresponding contribution to the energy of that state of order
$R_\infty\alpha^4$. Results of numerical calculation of the relativistic
corrections at this order are presented in Tables \ref{H2plus_a6}
and \ref{HDplus_a6}. For the transition frequency this adiabatic approach
provides about 3 significant digits.

\begin{table}[t]
\begin{center}
\caption{Relativistic corrections of $R_\infty m\alpha^4$ order
(in units $c^4\!\times\!(1\>\mbox{a.u.})$), $\mbox{H}_2^+$.}
\label{H2plus_a6}
\begin{tabular}{c@{\hspace{5mm}}ccccc}
\hline\hline
\vrule width0pt height 11pt
 & $v=0$ & $v=1$ & $v=2$ & $v=3$ & $v=4$ \\
\hline
\vrule width0pt height 11pt
$L\!=\!0$ & $-$0.042097 & $-$0.042908 & $-$0.043786 & $-$0.044732 & $-$0.045729 \\
$L\!=\!1$ & $-$0.042100 & $-$0.042912 & $-$0.043792 & $-$0.044740 & $-$0.045738 \\
$L\!=\!2$ & $-$0.042107 & $-$0.042922 & $-$0.043805 & $-$0.044757 & $-$0.045756 \\
$L\!=\!3$ & $-$0.042117 & $-$0.042938 & $-$0.043825 & $-$0.044782 & $-$0.045783 \\
$L\!=\!4$ & $-$0.042133 & $-$0.042959 & $-$0.043854 & $-$0.044818 & $-$0.045820 \\
\hline\hline
\end{tabular}
\end{center}
\end{table}

\begin{table}[t]
\begin{center}
\caption{Relativistic corrections of $R_\infty m\alpha^4$ order
(in units $c^4\!\times\!(1\>\mbox{a.u.})$), $\mbox{HD}^+$.}
\label{HDplus_a6}
\begin{tabular}{c@{\hspace{5mm}}ccccc}
\hline\hline
\vrule width0pt height 11pt
 & $v=0$ & $v=1$ & $v=2$ & $v=3$ & $v=4$ \\
\hline
\vrule width0pt height 11pt
$L\!=\!0$ & $-$0.042043 & $-$0.042738 & $-$0.043483 & $-$0.044278 & $-$0.045126 \\
$L\!=\!1$ & $-$0.042045 & $-$0.042741 & $-$0.043487 & $-$0.044284 & $-$0.045132 \\
$L\!=\!2$ & $-$0.042050 & $-$0.042748 & $-$0.043496 & $-$0.044295 & $-$0.045146 \\
$L\!=\!3$ & $-$0.042058 & $-$0.042759 & $-$0.043510 & $-$0.044312 & $-$0.045167 \\
$L\!=\!4$ & $-$0.042069 & $-$0.042773 & $-$0.043529 & $-$0.044336 & $-$0.045195 \\
\hline\hline
\end{tabular}
\end{center}
\end{table}

\section{Higher order radiative corrections.}

The electron ground state wave function to a good extent may be
approximated by $\psi_e(\mathbf{r}_e) =
C[\psi_{1s}(\mathbf{r}_1)+\psi_{1s}(\mathbf{r}_2)]$, where $\psi_{1s}$
is the hydrogen ground state wave function. So, the most important
$R_\infty\alpha^5$ order contributions can be evaluated using this
approximate wave function and the expressions:
\begin{equation}\label{a5}
\begin{array}{@{}l}
\displaystyle
E_{se}^{(5)} =
\alpha^5\sum_{i=1,2}
   \left\{
      \frac{Z_i^3}{m_e^2}
      \Biggl[
         -\ln^2{\frac{1}{(Z_i\alpha)^2}}
         +A_{61}\ln{\frac{1}{(Z_i\alpha)^2}}
         +A_{60}
      \Biggr]
      \left\langle\delta(\mathbf{r}_i)\right\rangle
   \right\},
\\[4mm]\displaystyle
E_{2loop}^{(5)} = \frac{\alpha^5}{\pi m_e^2}
   \left[
      B_{50}
   \right]
         \left\langle
            Z_1^2\delta(\mathbf{r}_1)\!+\!Z_2^2\delta(\mathbf{r}_2)
         \right\rangle,
\end{array}
\end{equation}
where the constants $A_{61}$, $A_{60}$, and $B_{50}$ are taken equal to
the constants of the $ 1s$ state of the hydrogen atom $A_{61}=5.419\dots$
\cite{Lazer60}, $A_{60}=-30.924\dots$ \cite{Pac93}, and $
B_{50}=-21.556\dots$ \cite{b50}. Worthy to say that the leading
contribution ($R_\infty\alpha^5\ln^2\alpha$) is exact.

\section{Results and Conclusion}

\begin{table}
\caption{Summary of contributions to the
$(v\!=\!0,L\!=\!0)\!\to\!(v'\!=\!1,L'\!=\!0)$
transition frequency (in MHz).} \label{summary}
\begin{center}
\begin{tabular}{l@{\hspace{12mm}}d@{\hspace{12mm}}d}
\hline\hline
\vrule height 10.5pt width 0pt depth 3.5pt
 & \mbox{H}_2^+ & \mbox{HD}^+ \\
\hline
\vrule height 10pt width 0pt
$\Delta E_{nr}$ & 65\,687\,511.0686    & 57\,349\,439.9717    \\
$\Delta E_{\alpha^2}$ &   1091.041(03) &          958.152(03) \\
$\Delta E_{\alpha^3}$ &   -276.544(02) &         -242.125(02) \\
$\Delta E_{\alpha^4}$ &     -1.997     &           -1.748     \\
$\Delta E_{\alpha^5}$ &      0.120(23) &            0.105(19) \\
\hline
\vrule height 10pt width 0pt
$\Delta E_{tot}$& 65\,688\,323.688(25) & 57\,350\,154.355(21)\\
\hline\hline
\end{tabular}
\end{center}
\end{table}

Various contributions to the frequency interval of the
fundamental transition are summarized in Table \ref{summary}.
Uncertainty in orders $R_\infty\alpha^2$ and $R_\infty\alpha^3$ are
primarily due to numerical uncertainty in calculation of leading order
terms like $\langle \mathbf{p}^4\rangle$ in the Breit-Pauli Hamiltonian,
or the Bethe logarithm, $\beta(L,v)$ (see Refs.~\cite{HD_BL,H2_BL} for the
details), and can be improved by more extensive calculations. We estimate
uncertainty due to finite size of nuclei as $\sim\!3\cdot10^{-4}$ MHz for
these transitions. So, the latter is so far negligible for the
ro-vibrational spectroscopy. For the contribution of order
$R_\infty\alpha^5$ the error bars are determined by the total contribution
of the terms with coefficients $A$ and $B$ in Eq.~(\ref{a5}).

Recently, the $(v,L)\!: (0,2)\!\to\!(4,3)$ ro-vibrational transition for the
$\mbox{HD}^+$ ion has been precisely measured in the experiment at the
D\"usseldorf university \cite{Sch07}. Comparison with theoretical
calculation demonstrates a very good agreement:
\[
\begin{array}{@{}r@{\,}l}
E_{\rm exp}&=214\,978\,560.6(5) \mbox{ MHz}
\\[0.5mm]
E_{\rm th}&=214\,978\,560.88(7) \mbox{ MHz}
\end{array}
\]

In conclusion, the relativistic corrections of order $R_\infty\alpha^4$
allow to reduce the relative accuracy of the fundamental transition
frequency in $\mbox{H}_2^+$ to about $3\cdot10^{-9}$ or 0.3 ppb. Further
improvement we expect to achieve by numerical estimate of coefficients
$A_{61}$, $A_{60}$, and $B_{50}$ from Eq.~(\ref{a5}) using the two-center
adiabatic (or external field) approximation. That may reduce the final
uncertainty by a factor of $5$-$10$ and the relative uncertainty to less
than $10^{-10}$. Eventually, it will make real the main goal of our
studies: improving of the $m_p/m_e$ mass ratio from the ro-vibrational
spectroscopy of $\mbox{H}_2^+$ and $\mbox{HD}^+$.

\section{Acknowledgement}

The author wants to express his gratitude to L.~Hilico and K.~Pachucki for
helpful remarks. The support of the Russian Foundation for Basic
Research under a grant No. 05-02-16618 is gratefully acknowledged.

\clearpage

\end{document}